\documentclass[aps,twocolumn,superscriptaddress,showpacs]{revtex4}
\usepackage{amssymb, amsmath}
\usepackage{color}
\usepackage{graphicx}

\begin{document}
\title{Josephson junction ratchet: effects of finite capacitances}
\author{Jakub Spiechowicz}
\affiliation{Institute of Physics, University of Silesia, 40-007 Katowice, Poland}
\author{Peter H{\"a}nggi}
\affiliation{Institute of Physics, University of Augsburg, 86135 Augsburg, Germany}
\affiliation{Nanosystems Initiative Munich, Schellingstr, 4, D-80799 M{\"u}nchen, Germany}
\author{Jerzy {\L}uczka}
\affiliation{Institute of Physics, University of Silesia, 40-007 Katowice, Poland}
\affiliation{Silesian Center for Education and Interdisciplinary Research, University of Silesia, 41-500 Chorz{\'o}w, Poland}
%
%
\begin{abstract}
We study transport in an asymmetric SQUID which is composed of a loop with three capacitively and resistively shunted Josephson junctions: two in series in one arm and the remaining one in the other arm. The loop is threaded by an external magnetic flux and the system is subjected to both a time-periodic and a constant current. We formulate the deterministic and, as well, the stochastic dynamics of the SQUID in terms of the Stewart-McCumber model and derive an equation for the phase difference across one arm, in which an effective periodic potential is of the ratchet type, i.e. its reflection symmetry is broken. In doing so, we extend and generalize earlier study by Zapata {\it et al.} [Phys. Rev. Lett. {\bf 77}, 2292 (1996)] and analyze directed transport in wide parameter regimes: covering the over-damped to moderate damping regime up to its fully under-damped regime. As a result we detect the intriguing features of a negative (differential) conductance, repeated voltage reversals, noise induced voltage reversals and solely thermal noise-induced ratchet currents. We identify a set of parameters for which the ratchet effect is most pronounced and show how the direction of transport can be controlled by tailoring the external magnetic flux.
\end{abstract}
\maketitle
\section{Introduction}
Josephson junctions are physical devices of prominent, wide spread scientific and practical use. Moreover, these can be used in testing the fundamentals of quantum mechanics and in studies for the many faces of chaotic complexity in classical physics. Scientists exploit them for a multitude of diverse theoretical and experimental studies. Applications in physics, electronics and other branches of engineering are well established: magnetometers, SQUIDs, superconducting qubits, RSFQ (rapid single flux quantum) circuitry - all use a Josephson junction as a primary building block. Here, we engineer a SQUID-device which is composed of three Josephson junctions and behaves as a physical ratchet system, i.e. a periodic structure which exhibits reflection-symmetry breaking \cite{rmphm,astumianPT,annalenhmn,reimann}.

A similar system was analyzed in Ref. \cite{zapata1996prl} for the over-damped case of the resistively shunted Josephson junctions. Here, we extend the study to include inertial effects by accounting for a finite capacitance (mass). This therefore leads to a modeling of the capacitively and resistively shunted case. In terms of classical mechanics, the former corresponds to the over-damped Brownian motion dynamics while the latter includes both finite dissipation and observable inertial effects. This extension is non-trivial because in the latter case the system allows for classical chaos. When the SQUID is driven by both a time-periodic and a constant current, it exhibits anomalous transport behavior including an absolute negative conductance in the linear response regime and negative static resistance in the nonlinear response regime.

This paper is organized as follows. In Sec. II we describe the circuit with three Josephson junctions and derive an equation which governs the dynamics of the studied system. Sec. III contains a detailed analysis of the deterministic transport processes occurring in our working model. In Sec. IV we study the role of thermal noise on the dynamics of the system. In Sec. V we seek the regime for which the ratchet effect arising in the device is most pronounced. In Sec. VI we propose the method of controlling the voltage direction by the external magnetic flux. Last but not least, Sec. VII provides a summary and conclusions. In the Appendix we derive an expression for the voltage across the SQUID.
\begin{figure}[b]
    \centering
    \includegraphics[width=0.89\linewidth]{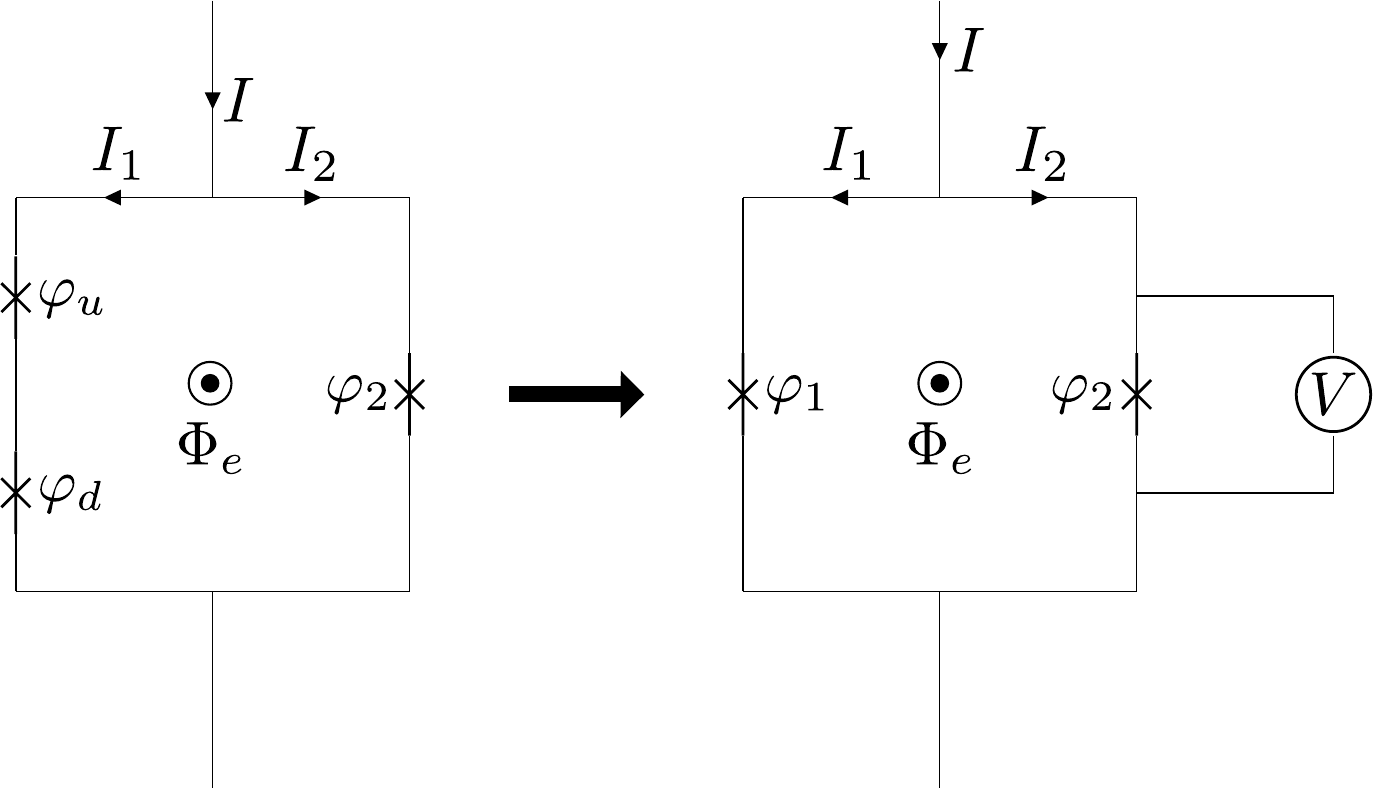}
    \caption{The asymmetric SQUID composed of three Josephson junctions and the equivalent circuit composed of two junctions, where the Josephson phase difference is $\varphi_1 = \varphi_u + \varphi_d$. The physical quantity of interest is the long-time average voltage $V$ across the SQUID which is expressed  by the relation: $V = \hbar \langle \dot{\varphi}_1 \rangle/2e = \hbar \langle \dot{\varphi}_2 \rangle/2e$, see Eq. (A1) in the Appendix.}
    \label{fig1}
\end{figure}
\section{Model}
We study transport properties of an experimental realization of the rocking ratchet mechanism in an asymmetric superconducting quantum interference device (SQUID) \cite{zapata1996prl, weiss2000, sterck2002, berger2004, sterck2005,sergey,sterck2009}. We analyze the current-voltage characteristics in the framework of the Stewart-McCumber theory \cite{stewart,mccumber}. The Stewart-McCumber model describes the semi-classical regime of a small Josephson junction for which a spatial dependence of characteristics can be neglected. Let us remind that in this theory the current $I(t)$ flowing through the junction is split into three components: the displacement current associated with its capacitance $C$, the normal Ohmic current due the finite resistance $R$ of the junction and the super-current of Cooper pairs characterized by the critical current $J$. Its explicit form reads
\begin{equation}
    \label{eq1}
    I(t) = C \dot{V}(t) + \frac{V(t)}{R} + J \sin{\varphi(t)},
\end{equation}
where $\varphi(t)$ is the phase difference between the macroscopic wave functions of the Cooper electrons in both sides of the junction, a dot denotes differentiation with respect to time $t$ and $V(t)$ is the voltage across the device which obeys the Josephson relation \cite{josephson}
\begin{equation}
    \label{eq2}
    V(t) = \frac{\hbar}{2e}\dot{\varphi}(t).
\end{equation}
If we insert (\ref{eq2}) into (\ref{eq1}) and include according to the fluctuation-dissipation relation \cite{kubo} the effect of a non-zero temperature $T>0$ by adding Johnson-Nyquist noise, the above Stewart-McCumber equation takes the form
\begin{equation}
    \label{eq2a}
    I(t) = \frac{\hbar}{2e} C \ddot{\varphi} + \frac{\hbar}{2e} \frac{1}{R} \dot{\varphi} + J \sin \varphi + \sqrt{\frac{2k_B T}{R}}\,\xi(t),
\end{equation}
where $\varphi \equiv \varphi(t)$, $k_B$ is the Boltzmann constant and thermal fluctuations are modeled by $\delta$-correlated Gaussian white noise $\xi(t)$ of zero mean and unit intensity
\begin{equation}
    \label{eq13}
    \langle \xi(t) \rangle = 0, \quad \langle \xi(t)\xi(s) \rangle = \delta(t-s).
\end{equation}
Following the proposal in Refs. \cite{zapata1996prl,sterck2005} we consider a SQUID ratchet which is composed of three Josephson junctions as sketched in Fig. \ref{fig1}. The loop contains two Josephson junctions in series in the left arm and one junction in the other arm. All elements are shunted with resistances ($R_u,R_d,R_2$) and corresponding capacitances ($C_u,C_d,C_2$). We safely can ignore the individual sub-gap resistances of the unshunted junctions, those being much larger than the shunt resistances. Moreover, the loop is pierced by an external magnetic flux $\Phi_e$. For each element in the left arm, exposed to the Kirchhoff left-arm current $I_1(t)$, we can write the Stewart-McCumber relation
\begin{subequations}
	\label{eq3}
	\begin{align}
	 I_1(t) &= \frac{\hbar}{2e} C_u \ddot{\varphi}_u + \frac{\hbar}{2e} \frac{1}{R_u} \dot{\varphi}_u + J_u \sin{\varphi_u} +\sqrt{\frac{2k_B T}{R_u}}\,\xi_u(t), \\
    I_1(t) &= \frac{\hbar}{2e} C_d \ddot{\varphi}_d + \frac{\hbar}{2e} \frac{1}{R_d} \dot{\varphi}_d + J_d \sin{\varphi_d} + \sqrt{\frac{2k_B T}{R_d}}\,\xi_d(t), 		
	\end{align}
\end{subequations}
where $\xi_u(t)$ and $\xi_d(t)$ are independent Gaussian white noises of the same statistics as in (\ref{eq13}). The processes $\xi_u(t)$ and $\xi_d(t)$ have to be independent to ensure the physically correct equilibrium Gibbs state.

Next, we consider the case when the two junctions in the left arm are identical, i.e. $J_u = J_d \equiv J_1, R_u = R_d \equiv R_1/2, C_u = C_d \equiv 2C_1$. Using these equal parameters we make us of the fact that ideally  the super-current in the \emph{left arm} is conserved.
Therefore, we find that the realization of the two phase solutions are synchronous in absence of the two noise terms for same initial conditions and temperature $T=0$. This singles out the unique and equal phases $\varphi_u=\varphi_d$. It implies that a solution of (5a) also obeys (5b) with same imposed left arm current $I_1(t)$\cite{zapata1996prl,sterck2005}. The Kirchhoff law remains valid also in presence of current noise noise with the identical (now random) left arm current $I_1(t)$. Because of the additional inhomogeneous Nyquist current noise term in each junction, however, the two solutions generally  stay no longer perfectly synchronized. Assuming small noise intensities for the  two thermal independent Gaussian noise sources of equal strength we approximate the phases as being synchronized nevertheless, i.e.   $\varphi_u=\varphi_d=\varphi_1/2$, with $\varphi_1 \equiv \varphi_u + \varphi_d$
Taking half of each relation in (\ref{eq3}a) and (\ref{eq3}b) and adding gives for the stochastic current $I_1(t)$ the result
\begin{align}
	\label{eq4}
	I_1(t) &= \frac{\hbar}{2e} C_1 \frac{d^2}{dt^2} (\varphi_u + \varphi_d) + \frac{\hbar}{2e} \frac{1}{R_1} \frac{d}{dt}(\varphi_u + \varphi_d) \nonumber \\
	&+ J_1\sin{\left( \frac{\varphi_u + \varphi_d}{2}\right )} \cos\left( \frac{\varphi_u - \varphi_d}{2} \right) \nonumber \\
	&+ \sqrt{\frac{k_B T}{R_1}}\,\xi_u(t) + \sqrt{\frac{k_B T}{ R_1}}\,\xi_d(t).
\end{align}
With equal solutions $\varphi_u=\varphi_d$ this expression yields the Langevin equation
\begin{equation}
    \label{eq5}
    I_1(t) = \frac{\hbar}{2e} C_1 \ddot{\varphi}_1 + \frac{\hbar}{2e} \frac{1}{R_1} \dot{\varphi}_1 + J_1\sin{\left( \frac{\varphi_1}{2} \right)} + \sqrt{\frac{2k_B T}{R_1}}\,\xi_1(t).
\end{equation}
Here, we used the fact that the linear combination of two independent Gaussian white noises of intensities $D_u = k_BT/2R_u$ and $D_d = k_BT/2R_d$ gives again Gaussian white noise with the total intensity described by $D_1=D_u+D_d=2k_BT/R_1$. Note that the stochastic process in (\ref{eq5}) amounts to a Johnson-Nyquist thermal noise for an overall shunt resistance $R_1 \equiv 2R_u = 2R_d$.

From the above analysis it follows that two identical junctions in series can be considered as one for which the supercurrent-phase relation assumes the form: $J_1\sin{\left(\varphi_1/2\right)}$ \cite{zapata1996prb,zapata1996prl,reimannpr}. This result was obtained in Ref.\cite{zapata1996prb} in the framework of the Ginzburg-Landau theory, cf. Eq. (23) therein. 

Let us discuss the above assumed synchronized phase approximation in presence of  small current noise in more detail. In the over-damped limit ($C_1=0$) this result agrees for identical junctions in the left arm as used for the three-junction SQUID rocking ratchet experiment investigated by Sterck \emph{et al.} \cite{sterck2005}, see Eqs. (4)-(7) therein. In reality, however, slight different junction parameters will physically lead to asynchronous phase variations in the two junctions in the left arm. Likewise, finite temperatures will, as indicated above,  destroy as well the perfect synchronous motion of the noisy solutions $\varphi_u= \varphi_d$, as assumed above at all times. However, the actual temperatures are experimentally very {\it small} \cite{sterck2002,sterck2005}. As it turns out, the physical ratchet effect for the average voltage emerging from this approximation remains itself  \emph{robust}. The latter has been verified before with simulations in the overdamped limit and also has been tested from experimental evidence in the corresponding low temperature limit. It was validated explicitly (i) numerically in \cite{zapata1996prl,sterck2005} and also (ii) experimentally for the three junction SQUID ratchet setup realized in the works \cite{sterck2005,sterck2009}.
Put differently, because we focus here on the Josephson voltage across the device, i.e. the \emph{average behavior} of the rate of change of the phase $\varphi_1$ but not on explicit stochastic values, the substitution of the $\cos[\left(\varphi_u - \varphi_d\right)/2]$-term by unity is justified in practice, as the corrections due to higher moments of the asynchronous phase difference can be safely neglected. In addition it must be kept in mind that the use of the Stewart-McCumber model is itself an approximation. Therefore, our theoretical predictions following from (\ref{eq5}) must be used as a guide towards "physical reality" for the experimenter rather than taken as granted without "error" \cite{zapata1996prl,reimannpr,sterck2005}.

For the junction in the right arm, the Stewart-McCumber equation reads
\begin{equation}
    \label{eq6}
I_2 (t) = \frac{\hbar}{2e} C_2 \ddot{\varphi}_2 + \frac{\hbar}{2e} \frac{1}{R_2} \dot{\varphi}_2 + J_2\sin \varphi_2 + \sqrt{\frac{2k_B T}{R_2}}\,\xi_2(t).
    \end{equation}
We next add the constraint for the phases in the loop threaded by the magnetic flux \cite{barone}
\begin{equation}
    \label{eq7}
    \varphi_2 - \varphi_1 = 2 \pi \frac{\Phi}{\Phi_0},
\end{equation}
where $\Phi_0 = h/2e$ is the flux quantum and the actual flux $\Phi$ is a sum of the external flux $\Phi_e$ and the flux due to the flow of currents
\begin{equation}
    \label{eq8}
    \Phi = \Phi_e + L i(t),
\end{equation}
where $L$ is the loop inductance and $i(t)$ is the \emph{circulating} current which tends to screen the magnetic flux. If the current is fed to the loop symmetrically then $i(t)=I_1(t) -I_2(t)$. An asymmetric case is presented in the Appendix. We consider the scenario when the second contribution is small, namely
\begin{equation}
    \label{eq9}
	|Li(t)| << \Phi_{0}.
\end{equation}
In this regime the internal flux increases monotonically with the external one and this operating mode is often called "dispersive" \cite{barone}. Then, from Eqs. (\ref{eq7})-(\ref{eq9}) we find
\begin{equation}
    \label{eq10}
    \varphi_2 = \varphi_1 + \tilde{\Phi}_e, \quad \tilde{\Phi}_e= 2\pi \frac{\Phi_{e}}{\Phi_0}.
\end{equation}
The total current $I(t)$ flowing through the SQUID is
\begin{equation}
    I(t) = I_1(t) + I_2(t).
	\end{equation}
We insert $I_1(t)$ and $I_2(t)$ from (\ref{eq5}) and (\ref{eq6}) and use (\ref{eq10}) to eliminate $\varphi_2$. The result is
\begin{align}
    \label{eq12}
	\frac{\hbar}{2e} C \ddot{\varphi_1} + \frac{\hbar}{2e} \frac{1}{R} \dot{\varphi}_1 &= -J_1\sin{\left( \frac{\varphi_1}{2} \right)} - J_2\sin{(\varphi_1 + \tilde{\Phi}_e)} \nonumber\\ &+ I(t) -
	\sqrt{\frac{2k_B T}{R}}\,\xi(t),
\end{align}
where $C = C_1 + C_2$ and $R^{-1} = R_1^{-1} + R_2^{-1}$.
The Gaussian white noise $\xi(t)$ is a linear combinations of $\xi_1(t)$ and $\xi_2(t)$ and has the same statistics as in (\ref{eq13}), cf. the similar transformation from (\ref{eq4}) to (\ref{eq5}).

Let the device be driven by an additional external current $I(t)$ which is composed of the static DC bias $I_0$ and the AC driving of amplitude $A$ and angular frequency $\Omega$, i.e.
\begin{equation}
    \label{eq14}
    I(t) = I_0 + A\cos(\Omega t).
\end{equation}
The mean value over the period $2\pi/\Omega$ is constant, $\langle I(t)\rangle = I_0$. As a consequence we obtain that
\begin{equation}
    \label{eq15}
    \frac{d \langle I(t)\rangle}{dt} = \frac{d \langle I_1(t)\rangle}{dt} + \frac{d \langle I_2(t)\rangle}{dt} = 0.
\end{equation}
In the Appendix, we show that in this case the voltage $V$ across the SQUID, averaged over the period of the AC current, is given by the relation
\begin{equation}
    \label{eq16}
	V = \frac{\hbar}{2e} \langle \dot{\varphi}_1 \rangle,
\end{equation}
where $\varphi_1$ is a solution of (\ref{eq12}) and $\langle \cdot \rangle$ denotes a temporal average over one period of the AC current.

\subsection{Going to a dimensionless formulation}
\begin{figure}[t]
    \centering
    \includegraphics[width=0.89\linewidth]{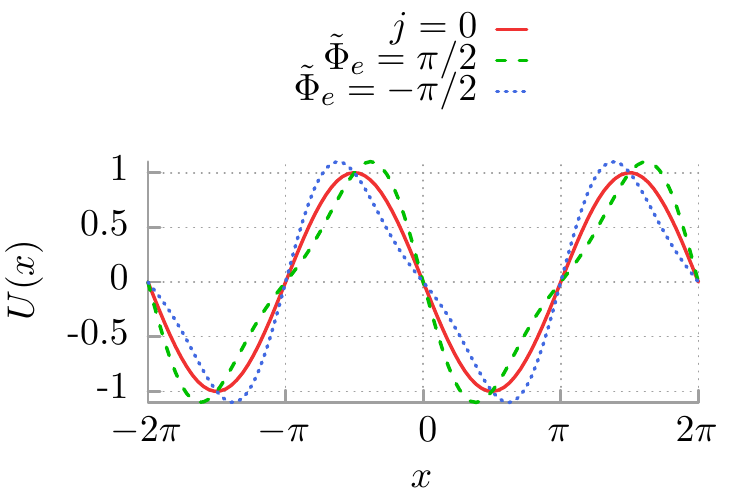}
    \caption{The symmetric potential $U(x) = -\sin(x)$ for $j=0$ (solid red line) is depicted in comparison with the ratchet potential given by (\ref{eq19}) for $j=1/2$ and two values of the external magnetic flux $\tilde\Phi_e = \pi/2$ (dashed green line) and $\tilde\Phi_e = -\pi/2$ (dotted blue line).}
    \label{fig2}
\end{figure}

We next transform (\ref{eq12}) into its dimensionless form. This can be achieved in several ways. It is known \cite{kautz} that for such a system there are four characteristic frequencies: plasma frequency $\omega_p^2 = 2eJ_1/\hbar C$, the characteristic frequency of the junction $\omega_c = 2eR J_1/\hbar$, the frequency $\omega_r =1/RC$ related to the relaxation time and the frequency $\Omega$ of the AC current. There are three independent characteristic time scales related to these frequencies (note that $\omega_p^2=\omega_c \omega_r$). Here, we follow \cite{zapata1996prl} and define the new phase $x$ and the dimensionless time $\hat{t}$ as
\begin{equation}
    \label{eq17}
     x = \frac{\varphi + \pi}{2}, \quad \hat{t} = \frac{t}{\tau_c}, \quad \tau_c = \frac{\hbar}{eRJ_1}.
\end{equation}
The corresponding dimensionless form of (\ref{eq12}) reads
\begin{equation}
    \label{eq18}
    \tilde C \ddot{x}(\hat{t}) + \dot{x}(\hat{t}) = -U'(x(\hat{t})) + F + a\cos(\omega \hat{t}) + \sqrt{2D}\,\hat{\xi}(\hat{t}),
\end{equation}
where the dot and prime denotes a differentiation over the dimensionless time $\hat t$ and the phase $x$, respectively. We introduced a spatially periodic potential $U(x)$ of period $2\pi$ of the following form \cite{zapata1996prl}
\begin{equation}
    \label{eq19}
    U(x) = - \sin(x) - \frac{j}{2} \sin(2x + \tilde\Phi_e - \pi/2).
\end{equation}
This potential is reflection-symmetric if there exists $x_0$ such that $U(x_0+x)=U(x_0-x)$ for any $x$. If $j \neq 0$, it is generally asymmetric and its reflection symmetry is broken. We classify this characteristics as a ratchet potential. However, even for $j \neq 0$ there are certain values of the external flux $\tilde \Phi_e$ for which it is is still symmetric. The dimensionless capacitance $\tilde C$ is the ratio between two characteristic time scales $\tilde C = \tau_r/\tau_c$, where the relaxation time is $\tau_r = RC$. Other re-scaled parameters are $j = J_2/J_1$, $F = I_0/J_1$, $a = A/J_1$ and $\omega = \Omega\tau_c$. It is worth to note that the noise intensity $D = e k_B T/\hbar J_1$ is the quotient of the thermal energy and the Josephson coupling energy. The re-scaled Gaussian white noise is of vanishing mean and the auto-correlation function \mbox{$\langle \hat{\xi}(\hat{t})\hat{\xi}(\hat{s}) \rangle = \delta(\hat{t} - \hat{s})$}. Hereafter, we will use only dimensionless variables and shall omit the 'hat' notation in all quantities appearing in (\ref{eq18}). In Fig. \ref{fig2}, the ratchet potential $U(x)$ is shown for $j=1/2$ and two values of the external magnetic flux $\tilde \Phi_e = \pi/2$ (positive "polarity") and $\tilde \Phi_e = -\pi/2$ (negative "polarity"). The symmetric potential for $j=0$ is also depicted. We would like to add that (\ref{eq18}) has a mechanical interpretation: it is identical to the Langevin equation of a classical Brownian particle of mass $m=\tilde C$ (i) moving in spatially periodic ratchet potential $U(x)$, (ii) being rocked by an unbiased harmonic force $a\cos(\omega t)$ and (iii) exposed to a static force $F$. In this mechanical framework the phase $x$ and the voltage $V$ translates to the space coordinate and the velocity of the Brownian particle, respectively.

The most important characteristic of transport behavior of the SQUID is the current-voltage curve in the stationary regime. The voltage (\ref{eq16}) or its dimensionless counterpart $\langle v \rangle = \langle \dot{x} \rangle$ is determined by (\ref{eq18}). In the long time limit, it takes the form of a Fourier series over all harmonics \cite{jung1993}, namely,
\begin{equation}
	\label{eq20}
	\lim_{t\to\infty} \langle {\dot x(t)} \rangle = \langle v \rangle + v_{\omega}(t) + v_{2\omega}(t) + \dots,
\end{equation}
where $\langle v \rangle $ is a DC (time-independent) component and $v_{n \omega}(t)$ are time-periodic functions of zero average over a basic period $2\pi/\omega$. In this case the DC component $\langle v \rangle$ is obtained after averaging over both the temporal period of the driving and the corresponding ensemble \cite{jung1993}
\begin{equation}
	\label{eq21}
	\langle v \rangle = \lim_{t\to\infty} \frac{\omega}{2\pi} \int_{t}^{t+2\pi/\omega} \mathbb{E}[\dot{x}(s)] \, ds,
\end{equation}
where $\mathbb{E}[\dot{x}(s)]$ denotes an average over initial conditions and all realizations of the thermal noise. The actual stationary voltage is then given as
\begin{equation}
    \label{eq22}
    V = R J_1 \langle v \rangle.
\end{equation}
Because the SQUID is driven by the external current (\ref{eq14}), the system is far away from thermal equilibrium and a time-dependent non-equilibrium state is reached in the long time limit. The key ingredient for the occurrence of directed transport $\langle v \rangle \neq 0$ is the symmetry breaking. It is the case when the DC current $F \neq 0$ or the reflection symmetry of the potential $U(x)$ is broken.
\begin{figure}[t]
    \centering
    \includegraphics[width=0.85\linewidth]{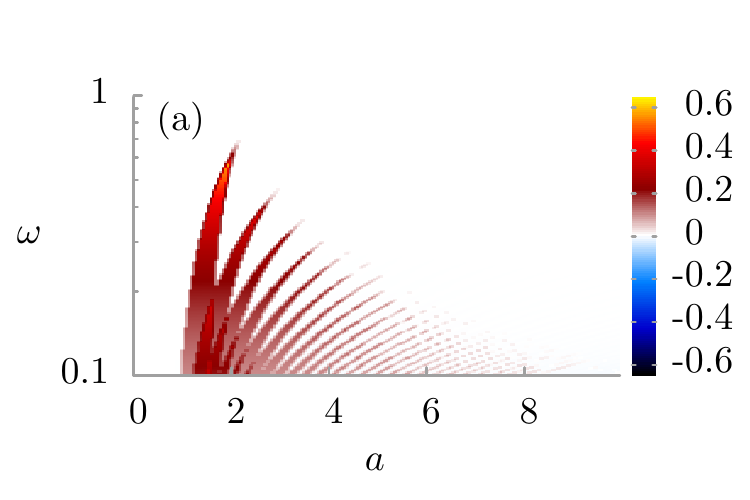} \\
    \includegraphics[width=0.85\linewidth]{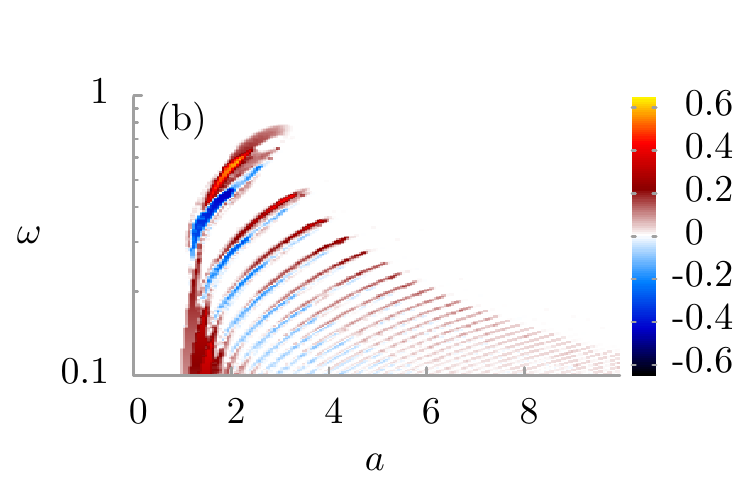} \\
    \includegraphics[width=0.85\linewidth]{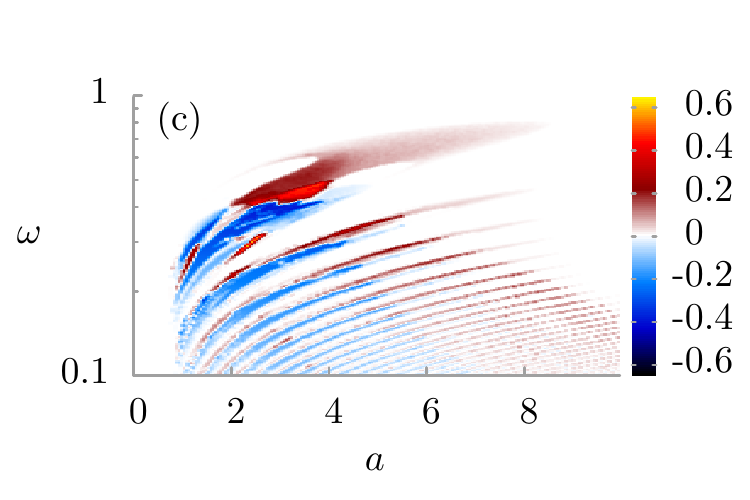}
    \caption{The deterministic transport behavior as a function of AC driving strength $a$ and its angular frequency $\omega$ of the dynamics in (\ref{eq18}) within three distinct regimes: (a) over-damped regime ($\tilde C = 0.2$), (b) moderate damping regime ($\tilde C = 2$) and (c) under-damped regime ($\tilde C = 7$). The average voltage $\langle v \rangle$ is presented for vanishing bias $F = 0$ and thermal noise $D = 0$. The periodic potential $U(x)$ has a positive "polarity": $j=1/2$ and $\tilde \Phi_e=\pi/2$.}
    \label{fig3}
\end{figure}
\section{Deterministic dynamics}
First, let us consider the corresponding deterministic version of the Langevin equation (\ref{eq18}), i.e. we set formally $D=0$. This is not the manifest realistic physical situation as thermal noise is present as well. However, it can help to understand general properties of the system. When $D=0$, (\ref{eq18}) is equivalent to a system of three autonomous differential equations of the first order and the phase space is three dimensional. It is a minimal dimension for chaotic behavior to occur. Indeed, periodic, quasi periodic and chaotic trajectories can be detected. A rough classification can be made into locked states in which the motion of $x$ is bounded to a few spatial periods and running states in which it is unlimited in space of $x$. The latter are crucial for the occurrence of the deterministic transport. For some regimes, ergodicity is broken and the systematic non-zero voltage emerges with its sign depending on the choice of selected initial conditions.
However, in the presence of small noise the system typically becomes ergodic and transitions between possibly coexisting deterministic disjoint attractors are probable. In particular, this give rise to diffusive directed transport.

In order to obtain the relevant transport characteristics we have to resort to comprehensive numerical simulations of driven Langevin dynamics. We integrated (\ref{eq18}) by employing a weak version of the stochastic second order predictor corrector algorithm \cite{platen} with a time step typically set to about $10^{-3} \cdot 2\pi/\omega$. Since (\ref{eq18}) is a second-order differential equation, we have to specify two initial conditions $x(0)$ and $\dot{x}(0)$. Moreover, because for some regimes the system may be non ergodic in order to avoid the dependence of the presented results on the specific selection of initial conditions we have chosen phases $x(0)$ and dimensionless voltages $\dot{x}(0)$ equally distributed over interval $[0, 2\pi]$ and $[-2,2]$, respectively. All quantities of interest were ensemble-averaged over $10^3 - 10^4$ different trajectories which evolved over $10^3 - 10^4$ periods of the external AC driving. Numerical calculations were done by use of a CUDA environment implemented on a modern desktop GPU. This scheme allowed for a speed-up of a factor of the order $10^3$ times as compared to a common present-day CPU method \cite{januszewski2009}.
\begin{figure*}[t]
 \centering
    \includegraphics[width=0.37\linewidth]{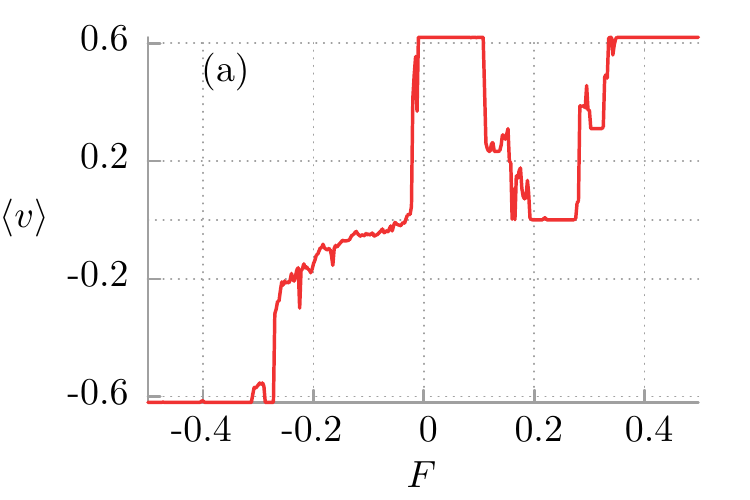}
    \includegraphics[width=0.37\linewidth]{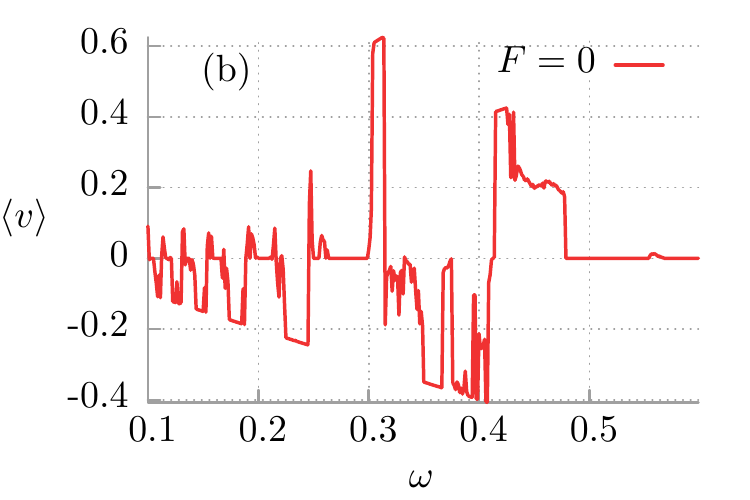} \\
    \includegraphics[width=0.37\linewidth]{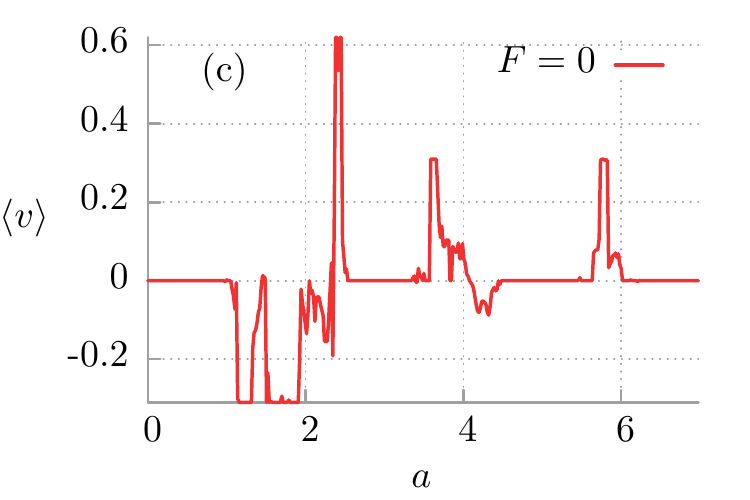}
    \includegraphics[width=0.37\linewidth]{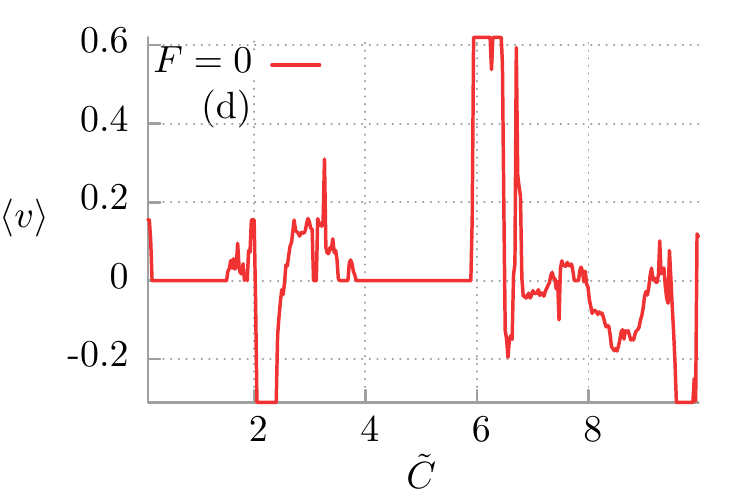}
    \caption{Representative transport characteristics of the rocked SQUID in the deterministic regime ($D = 0$) and for the potential $U(x)$ with a positive "polarity": $j=1/2$ and $\tilde \Phi_e=\pi/2$. Panel (a): current-voltage curve for driving strength $a = 2.4$, driving angular frequency $\omega = 0.31$ and capacitance $\tilde C = 6.31$. Panel (b): the dependence of the average voltage $\langle v \rangle$ on the AC-driving frequency $\omega$. Panel (c): Influence of the AC-driving amplitude $a$ on the DC voltage. Panel (d): the dependence of the average voltage $\langle v \rangle$ on the SQUID capacitance $\tilde C$. Remaining parameters in panels (b)-(d) are the same as in (a).}
    \label{fig4}
\end{figure*}
\subsection{General behavior}
The system described by (\ref{eq18}) possesses a 5-dimensional parameter space $\{\tilde C, a, \omega, F, D\}$. In this section we consider the deterministic case $D = 0$. Let us study the non-trivial ratchet effect by putting $F = 0$. Then, all forces on the right hand side of (\ref{eq18}) are zero on average: the mean potential force $-\langle U'(x)\rangle =0 $ on the interval $[x, x+2\pi]$ and the average AC driving
$\langle a \cos(\omega t) \rangle = 0$ on the time interval $[t, t+2\pi/\omega]$. If $\langle v \rangle \neq 0$, we detect the ratchet effect. Now, the parameter space $\{\tilde C, a, \omega\}$ is 3-dimensional and its exploration is tractable numerically with the currently available personal GPU computers. Depending on the value of the dimensionless capacitance $\tilde C$ the device can operate in three distinct regimes: over-damped ($\tilde C << 1$), moderate ($\tilde C \sim 1$) and under-damped ($\tilde C >> 1$). The first regime has been extensively studied in Refs. \cite{zapata1996prl,weiss2000,sterck2005,sergey,sterck2009}. In particular, it is known that in the deterministic case the average voltage $\langle v \rangle$ is almost quantized, displaying Shapiro-like steps in the current-voltage characteristic for the adiabatic and non-adiabatic AC driving frequencies $\omega$. As long as the potential $U(x)$ is asymmetric, generally $\langle v \rangle \neq 0$ with $F = 0$ \cite{bartussek1994,borromeo2002}. Since very fast positive and negative changes of the driving current cannot induce a non-zero average voltage, it is sufficient to limit our considerations to low and moderate AC driving frequencies $\omega$. We have performed scans of the parameter space: $\tilde C \times a \times \omega \in [0.1;10] \times [0;10] \times [0.1;1]$ at a resolution of 200 points per interval to determine the general behavior of the system. The results are depicted in Fig. \ref{fig3} for the positive "polarity" of the potential $U(x)$, i.e. for the external magnetic flux $\tilde \Phi_e = \pi/2$, cf. Fig. \ref{fig2}.

On all $(a, \omega)$ cuts, there occurs no ratchet effect for $a < 1$ and high driving frequencies $\omega$. The domains of non zero average voltage $\langle v \rangle$ have a striped structure. Although there is no obvious direct connection to chaotic properties of the system, we have found that for regimes where the ratchet effect is present a chaotic behavior is typically observed. For a fixed amplitude $a$, the ratchet behavior generally tends to disappear as the frequency $\omega$ grows. On the other hand, for a fixed frequency $\omega$, there is the optimal amplitude $a$ that maximizes the ratchet effect. The increase of the capacitance $\tilde C$ causes the appearance of regions for which the average voltage $\langle v \rangle$ reverses its sign. This should be contrasted with the over-damped regime in which the average voltage drop across the device is never negative for the potential with the positive polarity. Consequently, the capacitance $\tilde C$ of the device together with the amplitude $a$ and frequency $\omega$ of the AC-driving can serve as convenient parameters to manipulate the direction of transport processes occurring in the system (\ref{eq18}).
\subsection{Voltage {\it vs} DC current: Negative conductance}
Because the dynamics determined by (\ref{eq18}) is non-linear and the system is multidimensional, it should not come as surprise that the current-voltage curve is also non-linear and often depicts a non-monotonic function of the system parameters. Typically, the average voltage $\langle v \rangle$ is an increasing function of the DC-current $F$. This is true especially for large $F$. Such regimes correspond in the parameter space to normal, Ohmic like transport behavior. However, there are also regimes of anomalous transport exhibiting negative conductance \cite{kostur2006}: If the average voltage $\langle v \rangle$ is a decreasing function of the static bias $F$, the differential conductance
\begin{equation}
\label{eq23}
	\mu(F) = \left[\frac{dv(F)}{dF}\right]^{-1}
\end{equation}
can take negative values within some interval of $F$. Such a situation is depicted in panel (a) of Fig. \ref{fig4}. Clearly, there are several windows of the static current $F$ for which this effect is observed. It is worth to notice that this phenomenon is missing in the over-damped regime ($\tilde C \to 0$) or in the absence of the AC driving \cite{machura2007, speer2007epl}. Moreover, in panel (a) we show the ratchet effect: for $F=0$ the voltage is non-zero and for small negative DC current, $F<0$, the voltage is positive, $\langle v \rangle >0$. The latter phenomenon is named Absolute Negative Conductance (ANC) \cite{kostur2008}.
\begin{figure}[t]
    \centering
    \includegraphics[width=0.73\linewidth]{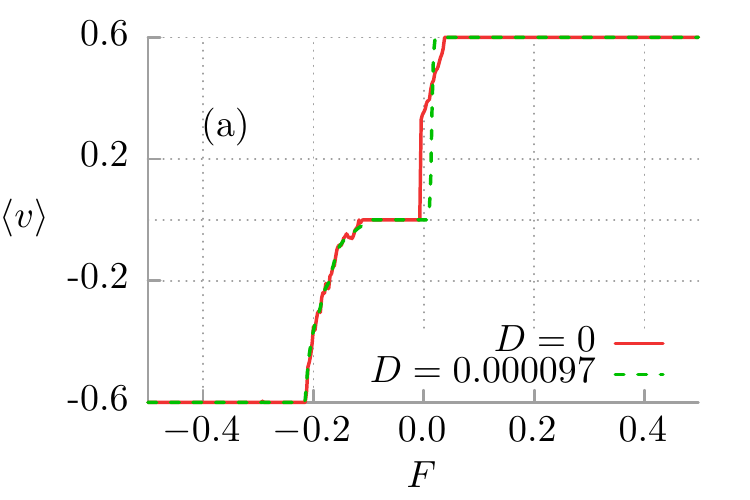} \\
    \includegraphics[width=0.73\linewidth]{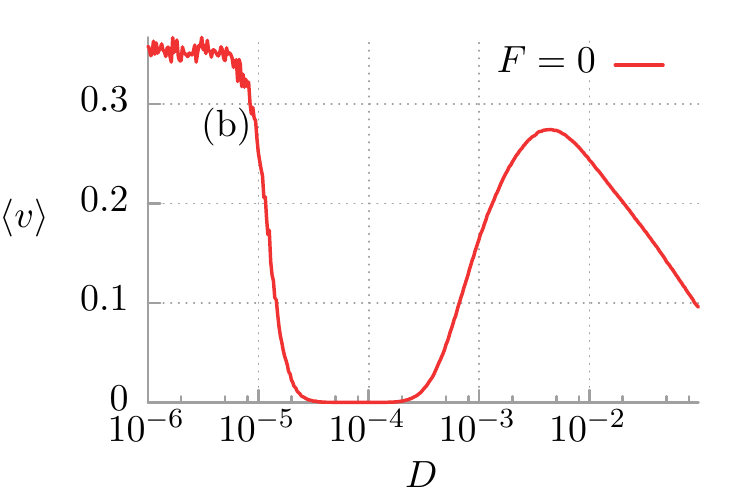}
    \caption{Destructive influence of thermal noise on the ratchet effect. Panel (a): The current-voltage characteristics is presented in the deterministic limit ($D = 0$, solid red line) and for the thermal noise driven case ($D = 9.7 \cdot 10^{-5}$, dashed green line) case. Panel (b): The dependence of the average voltage $\langle v \rangle$ on the thermal noise intensity $D$ for vanishing bias $F=0$. The remaining parameters are: $a = 1.9$, $\omega = 0.6$ and $\tilde C = 0.645$. The potential $U(x)$ has the positive "polarity": $j=1/2$ and $\tilde \Phi_e=\pi/2$.}
    \label{fig5}
\end{figure}
\subsection{Multiple voltage reversals}
According to the previous statement on the basis of general scans in the parameter space typical transport characteristics depicted in Fig. \ref{fig4} exhibit multiple reversals of the voltage $\langle v \rangle$ for the zero DC current, $F=0$. However, it should be stressed that this effect is not present in the over-damped regime ($\tilde C \to 0$) when for a fixed potential polarity the voltage has a fixed sign. The phenomenon of multiple voltage reversal \cite{jung1996,mateos2000,mateos2003,kostur2000} is most pronounced for moderate values of the amplitude $a$ and the frequency $\omega$ of the time-oscillating harmonic driving. In panel (b)-(d) we observe several local extrema and one global maximum of the voltage. For the increasing capacitance $\tilde C$ there are more regions in the parameter space for which this effect occurs. One can conveniently manipulate the direction of transport processes occurring in the system just by variation of its capacitance $\tilde C$, amplitude $a$ or frequency $\omega$.
\section{Role of thermal noise}
We can expect that thermal noise perturbs deterministic dynamics and can thus reduce or even destroy some deterministic effects. However, more interesting are the regimes for which it can enhance or induce new features for the system dynamics. We analyze the role of thermal fluctuations and discuss the influence of temperature on the stationary voltage. The regimes presented below are optimal in the sense that the effects are most pronounced for the illustrated parameter domains.
\begin{figure}[b]
    \centering
    \includegraphics[width=0.73\linewidth]{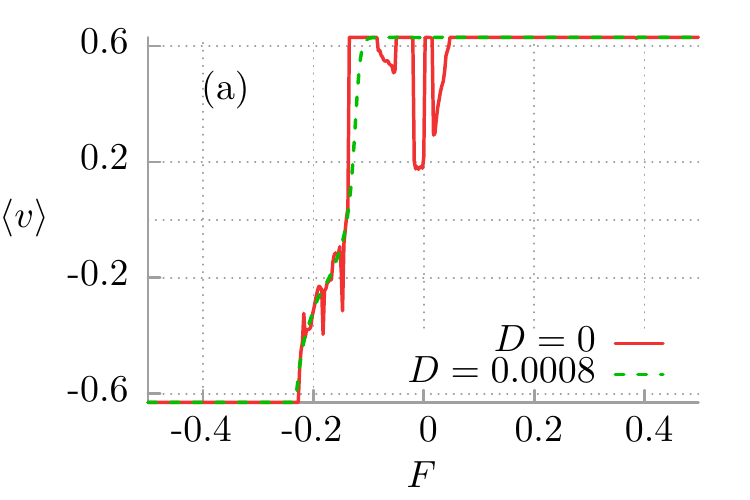} \\
    \includegraphics[width=0.73\linewidth]{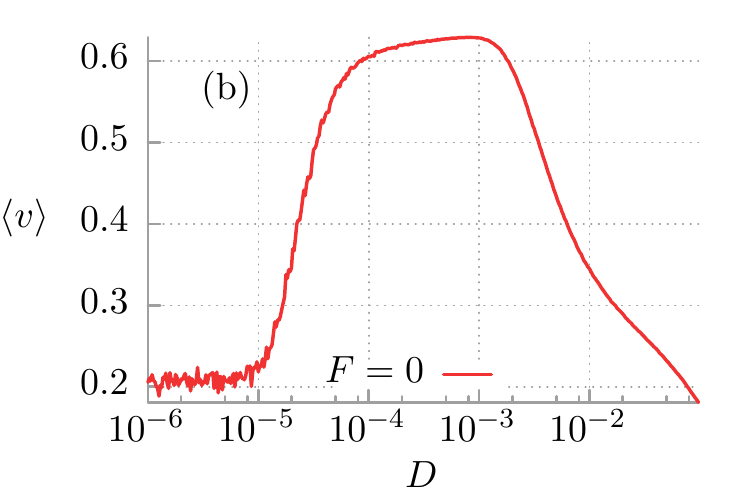}
    \caption{Constructive influence of thermal noise on the ratchet effect. Panel (a): The current-voltage characteristics is presented in the deterministic limit ($D = 0$, solid red line) and noise driven case ($D = 0.0008$, dashed green line) case. Panel (b): The dependence of the average voltage $\langle v \rangle$ on the thermal noise intensity $D$. The remaining parameters are: $a = 2.3$, $\omega = 0.63$, $\tilde C = 1.98$, $j=1/2$ and $\tilde \Phi_e=\pi/2$.}
    \label{fig6}
\end{figure}
\begin{figure*}[t]
    \centering
    \includegraphics[width=0.37\linewidth]{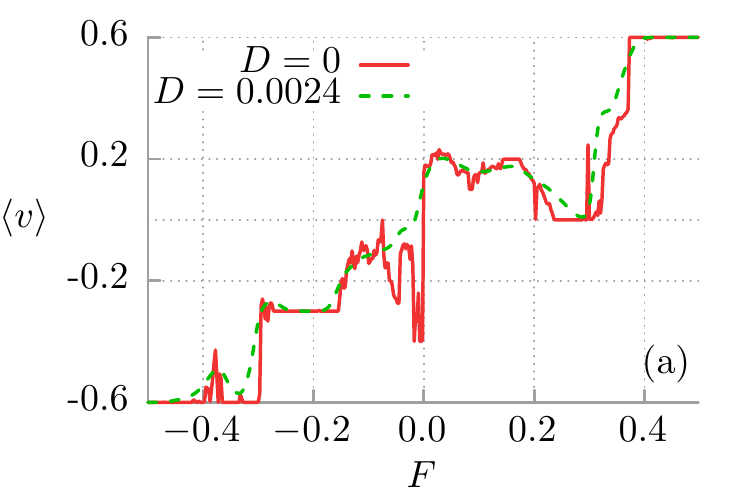}
    \includegraphics[width=0.37\linewidth]{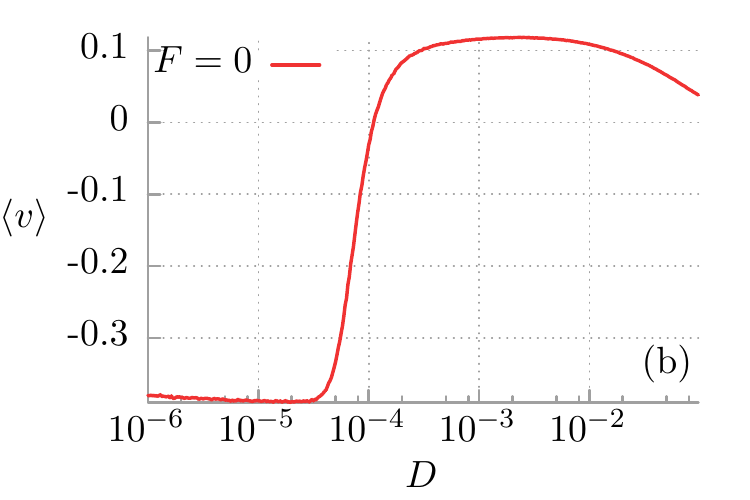} \\
    \includegraphics[width=0.37\linewidth]{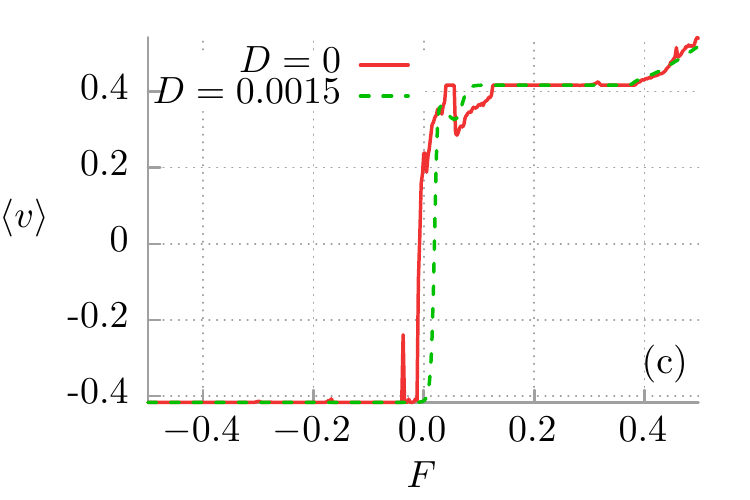}
    \includegraphics[width=0.37\linewidth]{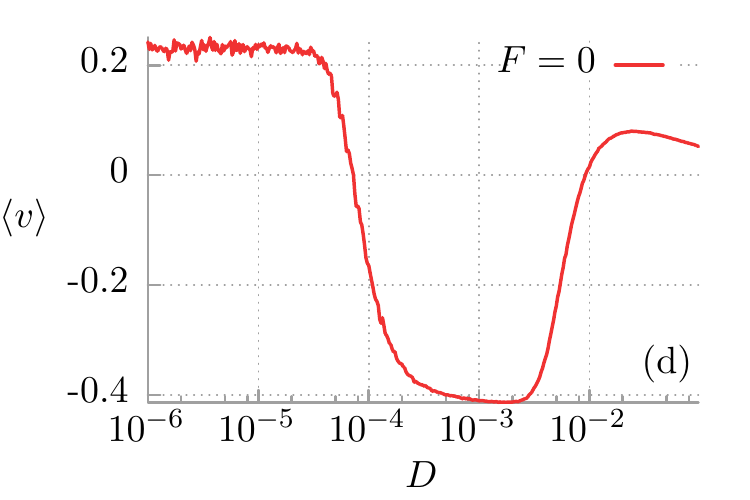}
    \caption{Destructive influence of thermal noise on the ratchet effect. In the panels (a) and (b) thermal noise may reverse the sign of the DC-voltage from negative to positive in comparison to the deterministic case. The remaining two panels (c) and (d) depict an opposite situation when the sign is shifted from positive to negative. Parameters for (a) and (b) read: $a = 1.7$, $\omega = 0.2$ and $\tilde C = 7.97$. For (c) and (d) they are as follows $a = 3.2$, $\omega = 0.417$, $\tilde C = 7$. The parameters of the potential $U(x)$ are: $j=1/2$ and $\tilde \Phi_e=\pi/2$.}
    \label{fig7}
\end{figure*}
\subsection{Destructive role of thermal fluctuations}
An example of a regime where thermal fluctuations play a destructive role is illustrated with Fig. \ref{fig5}. Panel (b) shows the dependence of the stationary average voltage $\langle v \rangle$ on the thermal noise intensity or temperature, $D \propto T$. A careful inspection of that figure reveals that indeed there is a window of temperature for which the DC-voltage is practically zero. One should also note that small increase of temperature causes a sharp reduction of the voltage $\langle v \rangle$ and therefore this phenomenon can be useful to trap the phase $x$ in one of the potential wells \cite{goldobin2007, sickinger2012, goldobin2013}. In panel (a) of the same figure we depict the current-voltage curves for the deterministic $D = 0$ and noisy $D = 9.7 \cdot 10^{-5}$ cases. Essentially, temperature plays a destructive role: there is no a ratchet effect for the noise intensity $D$ corresponding to the minimum of the curve in panel (b).
\begin{figure*}[t]
    \centering
    \includegraphics[width=0.37\linewidth]{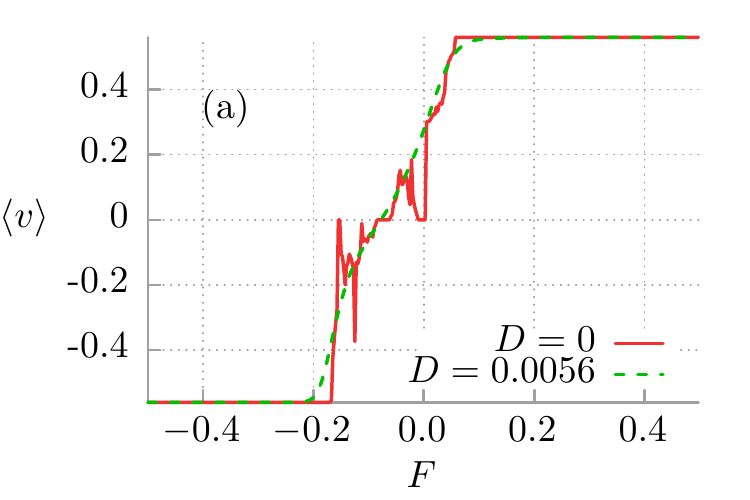}
    \includegraphics[width=0.37\linewidth]{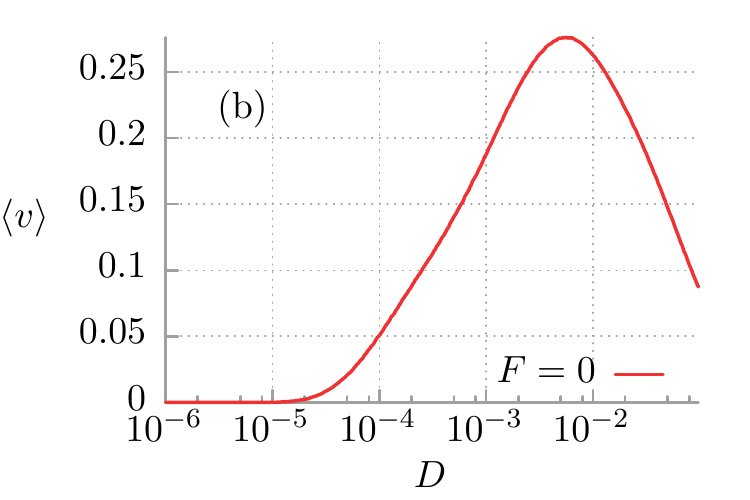}
    \caption{Noise induced ratchet effect. Panel (a) shows the current-voltage characteristic in this regime. Panel (b) depicts the dependence of the average voltage $\langle v \rangle$ on the thermal noise intensity $D$ in absence of a bias $F=0$. Other parameters are: $a = 1.8$, $\omega = 0.56$, $\tilde C = 1$, $j=1/2$ and $\tilde \Phi_e=\pi/2$.}
    \label{fig8}
\end{figure*}
\subsection{Constructive role of thermal fluctuations}
The opposite scenario occurs when thermal noise has a positive effect on relevant transport characteristics. It means that the voltage exhibits a maximum as a function of the thermal noise intensity $D$. In the mechanical framework it is equivalent to the situation when the mean first passage time for the particle to escape over the potential barrier is shortened by the increase of the thermal noise intensity $D$. This effect is exemplified in Fig. \ref{fig6}. Panel (b) shows the dependence of the average voltage $\langle v \rangle$ on temperature. Evidently, its increase causes an increase in the voltage. There is an optimal temperature, corresponding to $D \approx 0.0008$, for which the voltage $\langle v \rangle$ assumes a maximal value. This finding is confirmed in the current-voltage curve presented in the panel (a) of the same figure. Temperature plays a constructive role, the ratchet effect is strengthened by noise.
\begin{figure*}[t]
    \centering
    \includegraphics[width=0.3\linewidth]{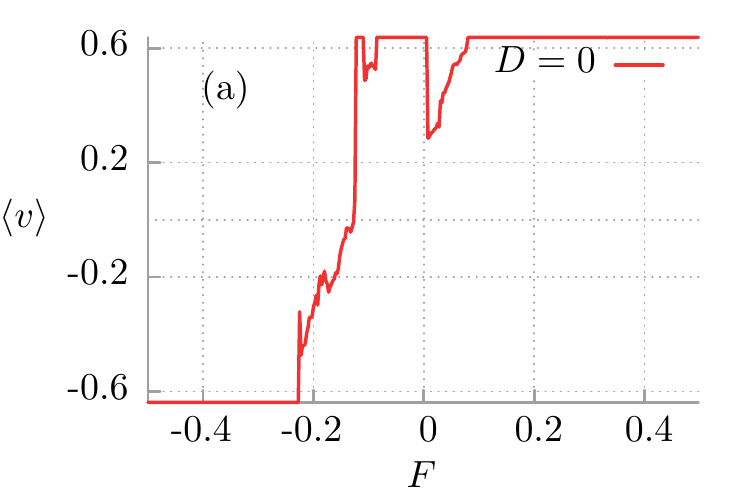}
    \includegraphics[width=0.3\linewidth]{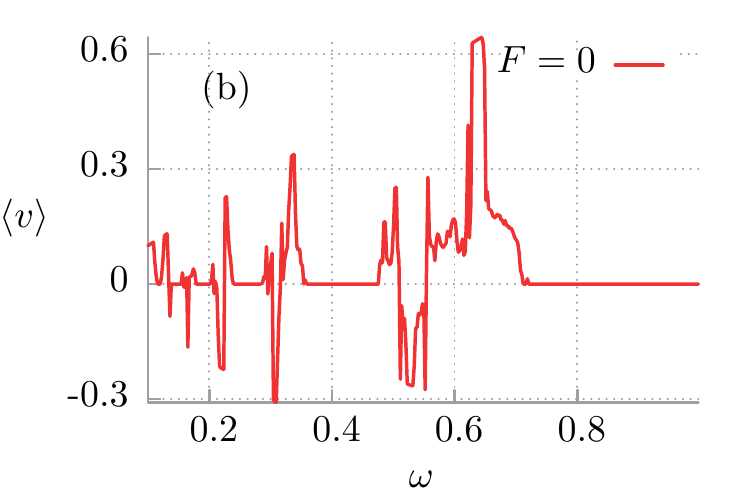}
    \includegraphics[width=0.3\linewidth]{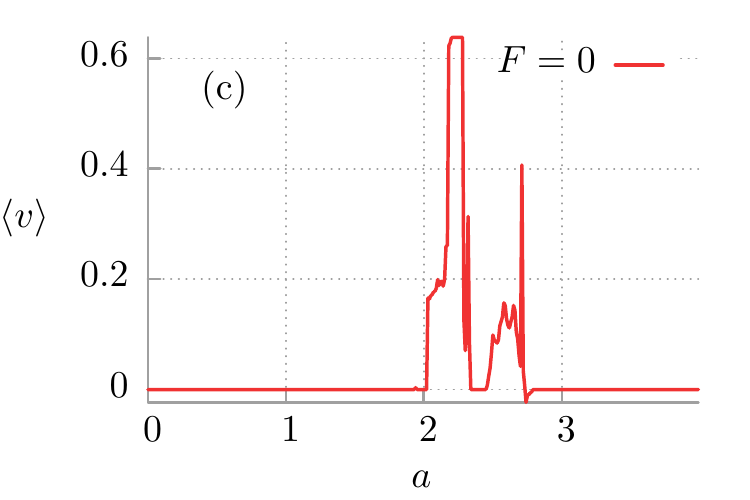} \\
    \includegraphics[width=0.3\linewidth]{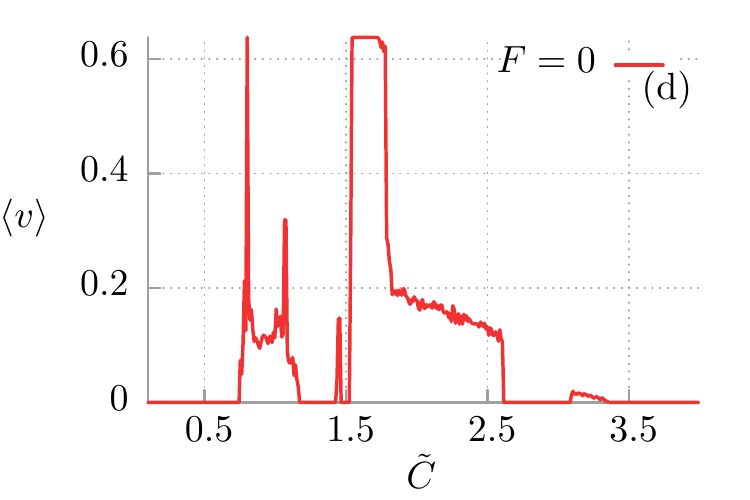}
    \includegraphics[width=0.3\linewidth]{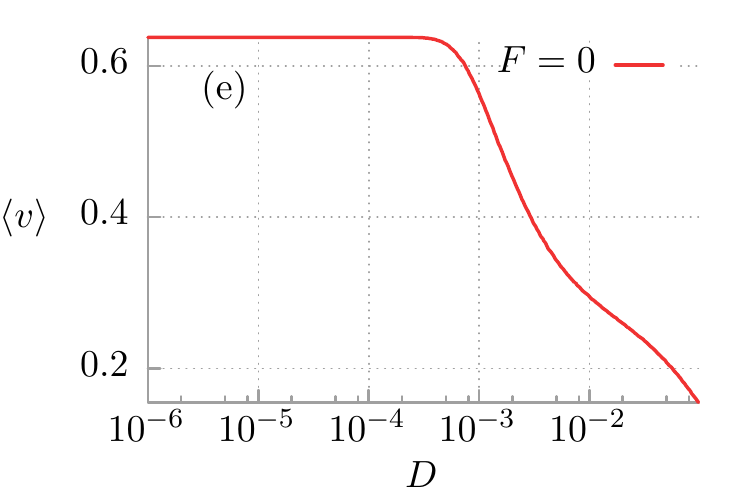}
    \caption{Optimal regime for the occurrence of the ratchet transport. The dependence of the average voltage $\langle v \rangle$ on the static DC-bias $F$, the angular frequency $\omega$, amplitude $a$, capacitance $\tilde C$ and thermal noise intensity $D$ is presented in panels (a)-(e). The chosen parameters are: $D=0$, $F = 0$, $a = 2.25$, $\omega = 0.638$, $\tilde C = 1.65$, $j=1/2$ and $\tilde \Phi_e=\pi/2$.}
    \label{fig9}
\end{figure*}
\subsection{Noise induced voltage reversals}
We have found a regime where thermal fluctuations are able to reverse the DC-voltage from positive to negative values and vice versa. Such regimes are illustrated in Fig. \ref{fig7}. In panel (b) of this figure the first situation is presented: the average voltage $\langle v \rangle$ is negative in the deterministic limit and increases with increasing temperature. There is a critical value of the thermal noise intensity $D$ for which the DC-voltage changes its sign and becomes positive for higher temperature. Panel (a) of the same figure shows the current-voltage curves corresponding to this regime. At this point it is worth to note that for this set of parameters the phenomenon of negative differential conductance is also detected. In particular, we can observe that this effect is robust with respect to small changes of the thermal noise intensity $D$. Panels (c) and (d) depict the opposite scenario: starting from low temperature the increase of $D$ changes the voltage from positive to negative values.
\subsection{Noise induced ratchet effect}
The next interesting phenomenon, which is activated by thermal fluctuations, is the noise induced ratchet effect. It corresponds to the situation when there is no directed transport in the deterministic regime $D = 0$ for vanishing static DC-bias $F = 0$ but it is observed when the thermal noise intensity is non-zero $D \neq 0$. Such a scenario is depicted in Fig. \ref{fig8}: The average voltage $\langle v \rangle$ vanishes for low thermal noise intensity $D$ and starts to increase with increasing temperature. There emerges also an optimal value of the thermal noise intensity $D \approx 0.0056$ for which the ratchet effect becomes most pronounced. This regime can be considered as a special case of a constructive influence of thermal noise on the ratchet phenomenon. Our finding is confirmed in the current-voltage curve which is presented in panel (a) of the same figure.
\section{Tailoring the ratchet current}
Modern personal GPU computers have given us opportunity to scan the parameter space of the system with high resolution in a reasonable time and therefore we were able to find a regime for which the ratchet effect is {\it globally} maximal, see in Fig. \ref{fig9}. All transport characteristics corresponding to this set of parameters are presented below. It turns out that the ratchet effect is optimal in the moderate capacitance regime $\tilde C \approx 1.65$, for the moderate amplitude $a \approx 2.25$ and the frequency $\omega \approx 0.638$ of the time-oscillating current. Moreover, there are several clearly indicated peaks in the dependence of the DC-voltage on the system parameters. The effect of thermal noise on the ratchet effect is destructive for this set of parameters. However, it is worth to note that this regime is temperature robust because the average voltage $\langle v \rangle$ starts to decrease significantly only for the thermal noise intensities higher than $D \approx 5 \cdot 10^{-4}$, cf. Fig. \ref{fig9}(e).
\section{Control of transport by external magnetic flux}
\begin{figure}[t]
	\centering
    \includegraphics[width=0.85\linewidth]{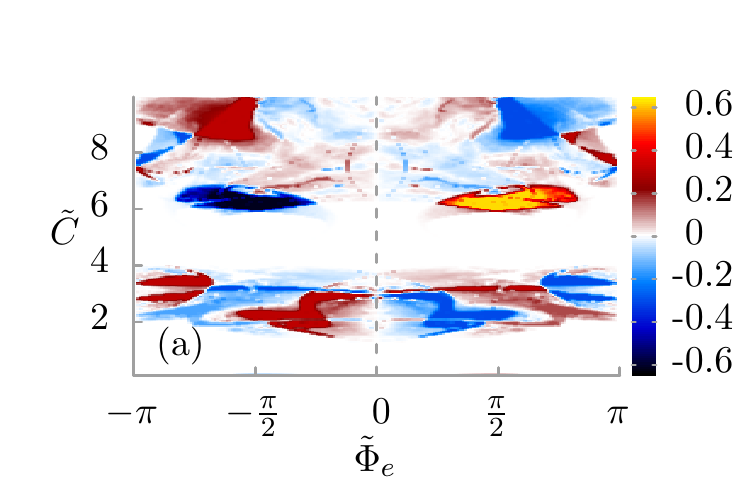}
    \includegraphics[width=0.85\linewidth]{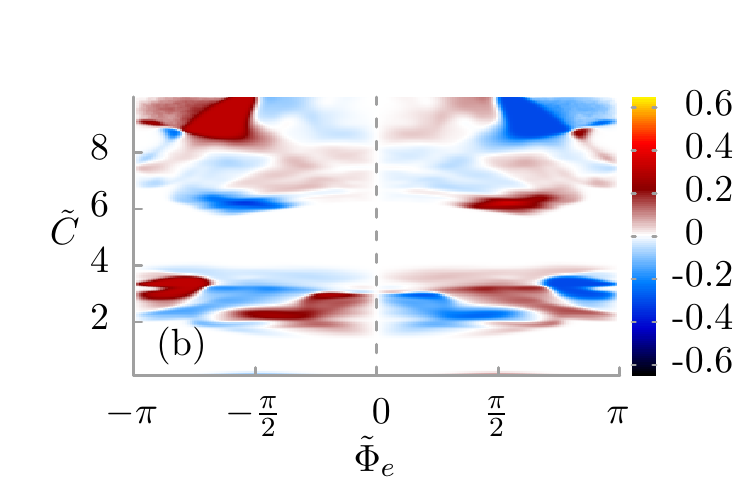}
    \caption{Voltage across the the rocked SQUID in the parameter plane $\{\tilde\Phi_e, \tilde C\}$. Upper panel: The deterministic case $D=0$. Bottom panel: The role of temperature $D=10^{-3}$. The DC current is absent, i.e. $F=0$. The remaining parameters are: $a = 2.4$, $\omega = 0.31, j=1/2$.}
    \label{fig10}
\end{figure}

Transport measured as the stationary DC-voltage can be controlled in a several ways. It seems that from the experimental point of view the simplest way is to vary a DC current $F$ or an external constant magnetic flux $\tilde \Phi_e$. We first consider the unbiased domain with $F=0$. In Fig. \ref{fig10} we depict how the DC-voltage behaves in the parameter plane $\{\tilde \Phi_e, \tilde C\}$ for two cases: $D=0$ (panel (a)) and $D= 10^{-3}$ (panel (b)). The most important feature of these plots is the symmetry with respect to the magnetic flux $\tilde\Phi_e$. For an arbitrary integer number $n$, the transformation $\tilde\Phi_e \to 2\pi n - \tilde\Phi_e$ reverses the polarity of the potential (\ref{eq19}) and as a consequence reverses also the voltage sign. The geometric structure of the domains in the depicted regime of the $\{\tilde \Phi_e, \tilde C\}$-variation is complex. There are islands of positive and negative voltage.

\begin{figure}[t]
 \centering
    \includegraphics[width=0.85\linewidth]{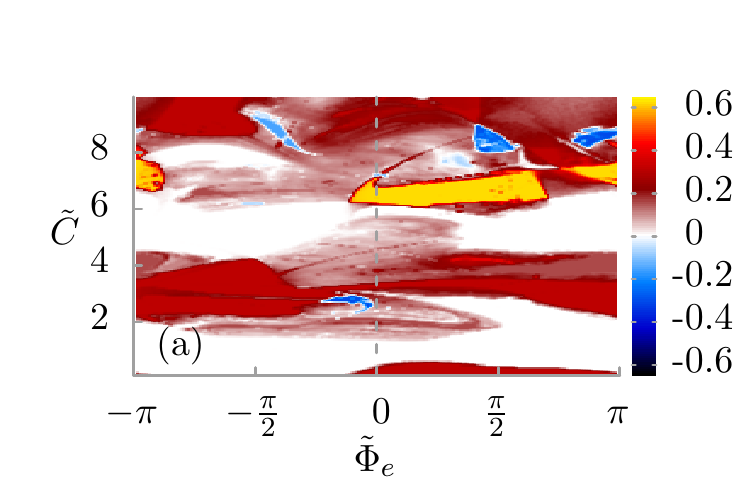}
    \includegraphics[width=0.85\linewidth]{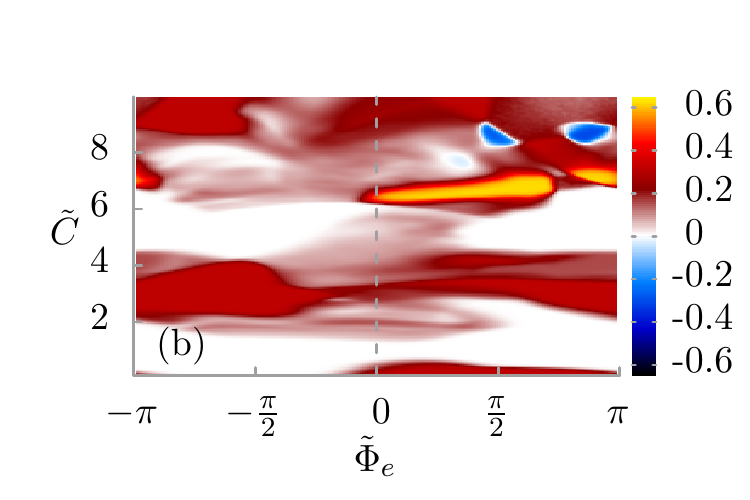}
    \caption{Voltage across the the rocked SQUID in the parameter plane $\{\Phi_e, \tilde C\}$. Upper panel: the deterministic case $D=0$. Bottom panel: influence of temperature $D=10^{-3}$. The DC current $F=0.1$. The remaining parameters are: $a = 2.4$, $\omega = 0.31, j = 1/2$.}
    \label{fig11}
\end{figure}

For the deterministic case ($D=0$) we reveal the refined structure. Some of these regions survive when the temperature is increased while others disappear. We detect a few robust regimes for which "islands" of non-zero voltage persist. It is seen that if the capacitance is fixed at the proper value the direction of transport can be changed by the magnetic field. In some regions, several voltage reversals can be obtained by use of this method. If the DC-current is applied, the above symmetry is destroyed. This case is shown in Fig. \ref{fig11}. However, there are still regimes where the magnetic field is a relevant control parameter for the direction of transport.
\section{Summary}
We analyzed the characteristics of the voltage across an asymmetric SQUID device composed of three capacitively and resistively shunted Josephson junctions which are threaded by a magnetic flux. We derived the evolution equation which governs the dynamics of the phase across the SQUID. The effective potential experienced by the phase displays a symmetry breaking in the form of the ratchet potential. Under the influence of an oscillating current source, the current-voltage characteristics yields the possibility to obtain a finite DC-voltage in presence of a vanishing DC-current, i.e. a ratchet effect is obtained. Within a tailored ranges of parameters, the same sign of the DC-voltage can be obtained regardless of the sign of the external DC-current.

With this comprehensive study we have taken into consideration the role of a finite capacitance of the SQUID. As a consequence, the resulting ratchet dynamics becomes rather rich, giving rise to features which are absent in the over-damped limit. With the help of the computational power of modern GPU computers we have identified a whole range of novel phenomena inherent for the ratchet current. These are a negative (differential) conductance, repeated DC-voltage reversals, noise induced DC-voltage reversals and particular forms of solely noise-induced ratchet features. For given tailored sets of parameters the ratchet voltage assumes optimal values.
Last but not least, we have been able to detect the set of parameters for which the ratchet effect is globally maximal and demonstrated how the direction of transport can be manipulated by tailoring the threading external magnetic flux.

The main goal of this work was the exploration and identification of parameter regimes for directed ratchet transport in realistic SQUID devices possessing {\it finite} capacitances. Such a study is of relevance for applications which make use of a generation and its control of the induced ratchet-voltages, their direction (sign), magnitude and their intrinsic sensitive dependence on system parameters.

Other transport quantifiers concerning the overall quality of the inertia-induced transport, such as the nature of the ratchet-voltage fluctuations (yielding in turn a diffusion dynamics of the phase across the SQUID), or the efficiency of the device \cite{machura2004, machura2005, machura2006,machura2010} have not been addressed here. Given the underlying complexity of the inertial ratchet dynamics these numerical studies are even more cumbersome than the presented ones.

Finally, an interesting question concerns the robustness of our results with respect to slightly different junction parameters in series; an assumed exact mathematically equality of parameters for two junctions is practically difficult to achieve. This issue has been addressed in the positive for the case of the over-damped regime \cite{zapata1996prl}, where it was found that the corresponding results remain robust. For the under-damped regime, the complexity of the problem becomes even more higher multi-dimensional and therefore this task is presently beyond the scope of this work. Nevertheless, those additional aspects are on our agenda when the corresponding cumbersome numerical investigations become technically more feasible.
\section*{Acknowledgments}
This work was supported in part by the MNiSW program ”Diamond Grant” (J. S.), NCN grant DEC-2013/09/B/ST3/01659 (J. {\L}.), and by a grant HA1517/-2 from the Deutsche Forschungsgemeinschaft (DFG) (P. H.). The authors also like to thank Peter Talkner for constructive discussions.
\appendix*
\section{Voltage across the asymmetric SQUID}
We demonstrate that in the "dispersive" operating mode of the SQUID, i.e. when the condition (\ref{eq9}) holds true, the averaged voltage developed across the SQUID can be expressed in the form:
\begin{equation}
    \label{A1}
	V = \frac{\hbar}{2e} \langle \dot{\varphi}_1 \rangle = \frac{\hbar}{2e} \langle \dot{\varphi}_2 \rangle.
\end{equation}
We follow the method presented in Ref. \cite{clarke} and consider an asymmetric junction configuration for which the total flux is
\begin{equation}
    \label{A2}
    \Phi = \Phi_e + \Phi_1 +\Phi_2 = \Phi_e + \mathcal L_1 I_1(t) - \mathcal L_2 I_2(t),
\end{equation}
where $\Phi_1$ and $\Phi_2$ are the fluxes produced by the currents $I_1(t)$ and $I_2(t)$, respectively. The coefficients $\mathcal L_1$ and $\mathcal L_2$ are related to the loop inductance via the relation
\begin{equation}
    \label{A3}
    L = \mathcal L_1 + \mathcal L_2
\end{equation}
The total voltage across the SQUID calculated along the left arm is the sum
\begin{equation}
    \label{A4}
	V = V_1 + L_1 \langle \dot{I}_1(t) \rangle + M \langle \dot{I}_2 (t) \rangle.
\end{equation}
Similarly, along the right side we have
\begin{equation}
    \label{A5}
	V = V_2 + L_2 \langle \dot{I}_2(t) \rangle + M \langle \dot{I}_1(t) \rangle,
\end{equation}
where $L_1$ and $L_2$ are self-inductances of the left and the right arm, respectively. Generally, they are different from $\mathcal L_i$, see Refs. \cite{fulton,clarke}. The mutual inductance between the two arms is $M$ and
\begin{equation}
    \label{A6}
	V_i = \frac{\hbar}{2e} \langle \dot{\varphi}_i \rangle, \quad i=1,2
\end{equation}
are the voltage drops across the left and right junctions, respectively. Now, we use (\ref{eq15}) to get
\begin{subequations}
	\label{A7}
	\begin{align}
	V &= V_1 + L_1 \langle \dot{I}_1(t) \rangle - M \langle \dot{I}_1(t) \rangle, \\
	V &= V_2 - L_2 \langle \dot{I}_1(t) \rangle + M \langle \dot{I}_1(t) \rangle.
	\end{align}
\end{subequations}
Adding both sides of these equations and utilizing (\ref{A6}) yields
\begin{equation}
    \label{A8}
	2 V = (\hbar/2e) (\langle \dot{\varphi}_1 \rangle + \langle \dot{\varphi}_2 \rangle) +(L_1-L_2) \langle \dot{I}_1(t)\rangle.
\end{equation}
By differentiation (\ref{eq10}) and making use of (\ref{eq15}) we obtain
\begin{align}
    \label{A9}
	\frac{\hbar}{2e}(\langle \dot{\varphi}_1 \rangle - \langle \dot{\varphi}_2 \rangle) &= \frac{2\pi}{\Phi_0} \dot{\Phi} = \frac{2\pi}{\Phi_0} \left( \mathcal L_2 \langle \dot{I}_2(t) \rangle - \mathcal L_1 \langle \dot{I}_1(t) \rangle \right ) \nonumber \\ &= - \frac{2\pi}{\Phi_0}( \mathcal L_1 + \mathcal L_2 ) \langle \dot{I}_1(t)\rangle.
\end{align}
From this equation we calculate $\langle \dot{I}_1(t) \rangle$ and insert it into (\ref{A8}) to obtain
\begin{equation}
    \label{A10}
	2 V = \left( \frac{\hbar}{2e} + \epsilon \right) \langle \dot{\varphi}_1 \rangle + \left( \frac{\hbar}{2e} - \epsilon \right)\langle \dot{\varphi}_2 \rangle,
\end{equation}
where
\begin{equation}
    \label{A11}
\epsilon = \frac{1}{L} (L_1 -L_2).
\end{equation}
Up to here we used (\ref{eq15}). From (\ref{eq9}) and (\ref{eq10}) it follows that $\langle \dot{\varphi}_1 \rangle = \langle \dot{\varphi}_2 \rangle$. In this case the result (\ref{A1}) follows directly from the result in (\ref{A10}).

\end{document}